\documentclass[12pt]{article}

\expandafter\let\csname equation*\endcsname\relax
\expandafter\let\csname endequation*\endcsname\relax
\usepackage{amsmath,amssymb,mathtools}
\usepackage{graphics}
\usepackage{epstopdf}
\usepackage{cite}
\usepackage[margin=0.8in]{geometry}

\usepackage[colorlinks,linkcolor=blue,urlcolor=blue,citecolor=blue]{hyperref}

\usepackage{cleveref}

\def\be{\begin{equation}}
\def\ee{\end{equation}}
\def\bea{\begin{eqnarray}}
\def\eea{\end{eqnarray}}

\usepackage{xcolor}
\definecolor{dgreen}{rgb}{0,0.7,0}

\begin{document}

\title{Autocorrelation functions and ergodicity in diffusion with stochastic resetting}

\author{Viktor Stojkoski$^{1,2}$, \\ Trifce Sandev$^{2,3,4}$,\\ Ljupco Kocarev$^{2,5}$ \\ Arnab Pal$^{6,\footnote{Corresponding author: \href{mailto:arnabpal@mail.tau.ac.il}{arnabpal@mail.tau.ac.il} }}$ \\ \\ \footnotesize
$^{1}$Faculty of Economics, Ss.~Cyril and Methodius University, 1000 Skopje, Macedonia \\
\footnotesize
$^{2}$Research Center for Computer Science and Information Technologies, Macedonian Academy of Sciences and Arts \\\footnotesize
$^{3}$Institute of Physics \& Astronomy, University of Potsdam \\\footnotesize
$^{4}$Institute of Physics, Faculty of Natural Sciences and Mathematics,
Ss Cyril and Methodius University\\\footnotesize
$^{5}$Faculty of Computer Science and Engineering, Ss. Cyril and Methodius University\\\footnotesize$^{6}$School of Chemistry, The Center for Physics and Chemistry of Living Systems, Tel Aviv University}
\date{\today}

\maketitle

\begin{abstract}
Diffusion with stochastic resetting is a paradigm of resetting processes. Standard renewal or master equation approach are typically used to study steady state and other transport properties such as average, mean squared displacement etc. What remains less explored is the two time point correlation functions whose evaluation is often daunting since it requires the implementation of the exact time dependent probability density functions of the resetting processes which are unknown for most of the problems. We adopt a different approach that allows us to write a stochastic solution in the level of a single trajectory undergoing resetting. Moments and the autocorrelation functions between any two times along the trajectory can then be computed directly using the laws of total expectation. Estimation of autocorrelation functions turns out to be pivotal for investigating the ergodic properties of various observables for this canonical model. In particular, we investigate two observables (i) sample mean which is widely used in economics and (ii) time-averaged-mean-squared-displacement (TAMSD) which is of acute interest in physics. We find that both diffusion and drift-diffusion processes are ergodic at the mean level unlike their reset-free counterparts. In contrast, resetting renders ergodicity breaking in the TAMSD while both the stochastic processes are ergodic when resetting is absent.  We quantify these behaviors with detailed analytical study and corroborate with extensive numerical simulations. The current study provides an important baseline that unifies two different approaches, used ubiquitously in economics and physics, for studying the ergodic properties in diffusion with resetting. We believe that our results can be verified in single particle experimental set-ups and thus have strong implications in the field of resetting.
\end{abstract}

\section{Introduction}
\label{introduction}
Resetting or restart can be a natural or man-made act. In chemical reactions, unbinding of an enzyme molecule from a substrate is often considered to be a restart event \cite{Restart-Biophysics1}. Similarly, dissociation phase of a protein from the DNA in the facilitated diffusion mechanism (where a protein searched for a target DNA) \cite{DNA} or the backtracking by RNA polymerases \cite{Restart-Bio-1} can be understood as restart events. On the other hand, in a data transmission system, reset is usually done in response to an error condition when it is impossible or undesirable for a processing activity to proceed and all error recovery mechanisms fail. Very recently, resetting has seen a surge of interest in statistical physics. Since the seminal work of Evans and Majumdar~\cite{EvansMajumdarPRL,Restart2} a substantial amount of research has been carried out on resetting and its various applications. This volume of work spans different fields starting from non-equilibrium transport properties \cite{transport1,transport2,Pal-potential,Pal-time-dep,powerlaw-diff,Durang,axel1,axel2,bodrova,CTRW1,CTRW2}, first-passage \cite{Levy-flight,Kirone,Levy,ReuveniPRL,PalReuveniPRL,PalReuveniBranching,cost,Chechkin,Belan,peclet,CV-c2,CV-suff} and search theory \cite{Restart-Biophysics4,Montanari,bressloff,HRS}, stochastic thermodynamics \cite{sth-0,sth-1,sth-2,sth-3}, extreme value statistics \cite{evs-1,evs-2} and stochastic resonance \cite{SR}. Along with the rapid progress in the theoretical front, resetting has also become an active playground for several experimental groups which have led to new understanding \cite{expt-1,expt-2,expt-3}. We refer to \cite{review} by Evans \textit{et al} which provides an overview of the current state of the problem.

Diffusion with resetting introduced by Evans and Majumdar is a canonical model in the field. According to this model, motion of a diffusing particle is interrupted at a rate $r$, and the particle is taken back to its initial position. Repeated such action curtails the detrimental trajectories of the diffusing particle and thus at a long time a non-equilibrium steady state is attained. This generic behavior was also identified for a particle diffusing in potential landscape \cite{Pal-potential,log-potential,jpa2020_rk}. Emergence of steady states turns out to be a robust feature of stochastic resetting as such observations were also made in a wide spectrum of other stochastic processes namely anomalous diffusion \cite{ana-1,ana-2}, scaled Brownian motion \cite{scaled}, geometric Brownian motion \cite{GBM}, random acceleration process \cite{RAP-1,RAP-2}, random walks in networks \cite{RW1,RW2}, continuous time random walk \cite{CTRW1,CTRW2}, L\'evy flights \cite{Levy-flight}, telegraphic process \cite{telegraphic}, and active particle systems \cite{RTP,RTP-2,RT}. Steady state properties were also studied in non-instantaneous resetting processes \cite{non-inst-0,non-inst-1,non-inst-2,non-inst-3,non-inst-4,non-inst-5}.

Besides steady states, other transport properties such as the average and mean squared displacement (MSD) of a particle described by a generic propagator were studied at all times by Mas\'{o}-Puigdellosas et al. \cite{axel1,axel2}. Bodrova \textit{et al} studied moments and MSD for scaled BM in the presence of resetting \cite{bodrova}. MSDs for stochastic processes with non-instantaneous resetting were studied by Bodrova \textit{et al} \cite{non-inst-3}. However, the two point time correlations are less known and to the best of our knowledge were studied by Oshanin and Majumdar \cite{oshanin} for simple diffusion using a renewal formalism. The renewal approach while very powerful and elegant requires complete information of the underlying process. In other words, to compute the autocorrelation functions with resetting, one requires the autocorrelation function of the underlying process. Somewhat similar, our approach uses the renewal property but in the single trajectory level instead of using the same in the ensemble level. This allows us
to write a stochastic solution (say in terms of position which is a random variable) for the resetting Langevin equation. Subsequently we compute the moments and autocorrelation functions applying the \textit{law of total expectation} to the random variable (e.g., position) directly.

Computation of autocorrelation functions serves as a bedrock for us to develop the second central aim of this work: testing the ergodic hypothesis in diffusion with stochastic resetting.
The ergodic hypothesis is a key analytical device in statistical physics which states that the time average and the expectation value of an observable are the same. 
Surprisingly enough, ergodic properties of resetting systems were not investigated until recently. In a recent work \cite{GBM}, we studied ergodicity of geometric Brownian motion subject to Poissonian resetting. We showed that the process is nonergodic in the level of \textit{sample mean} \cite{GBM}. 
This was realized in the longtime
behavior of the sample mean which resulted in
three regimes: i) a frozen state
regime, ii) an unstable annealed regime, and iii) a stable
annealed regime. These regimes are separated by a
self-averaging time period which depends strongly on the
resetting time density. While resetting does not affect the non-ergodicity of
the process in the first regime, we showed how resetting
non-trivially ramifies the self-averaging behavior leading
to either an unstable or a stable long-time sample average \cite{GBM}. In parallel, Wang \textit{et al} studied ergodic properties of fractional Brownian motion (FBM) and heterogeneous diffusion process (HDP) in the presence of resetting \cite{biorxiv}. However, they investigated the effects of resetting on the single trajectory based 
\textit{time-averaged-mean-squared-displacement }(TAMSD) and the ergodicity breaking parameter in the above-mentioned processes. The authors observed non-equivalence of the TAMSD and the mean-squared deviation for reset superdiffusive FBMs (only exceptions are reset subdiffusive FBMs), reset HDPs and for generalized process of reset HDP-FBM. Thus these two studies look into the ergodic properties for different reset processes, and crucially the quantifying observables (namely sample mean and TAMSD) were considered differently. 

The sample mean is of innate interest in economics, where it is traditionally applied for assessing the joint performance of a group of financial instruments and reducing the potential negative impact of random fluctuations on observed phenomena~\cite{peters2013ergodicity,stojkoski2019cooperation,stojkoski2021evolution}. In statistical physics a similar role is being played by the TAMSD. In particular, the TAMSD has been
implemented ubiquitously in single-particle-tracking experiments
and its characteristic features have been intensely developed
theoretically over the last years for a variety of non-reset
stochastic processes featuring anomalous diffusion, see the reviews \cite{TAMSD-review-1,TAMSD-review-2,TAMSD-review-3}. In this work, we unite both approaches and study the ergodic hypothesis using the canonical model of diffusion (with/without drift) in the presence of stochastic resetting as a testing ground. Our main objective here is to enrich the list of properties shown by both the observables: sample mean and the TAMSD employed in resetting dynamics of diffusion. In the absence of resetting, diffusion process (with or without drift) is known to be non-ergodic in the mean level, but we show that introduction of resetting renders ergodicity. On the other hand, the TAMSD is ergodic for both simple diffusion and drift-diffusion processes in the absence of resetting. We unveil a pronounced resetting induced non-ergodicity for TAMSD in both cases. Extensive numerical simulations support our analytical results. 

For brevity, we now outline the structure of the paper. In Section~\ref{sec:diffusiong-with-resetting} we present our approach for writing explicit solutions to diffusion with resetting on a single trajectory level. In the same section we discuss the properties of the Poissonian resetting time density. These properties are subsequently utilized in Section~\ref{sec:law-of-total-expectation} to quantify the moments and autocorrelation functions in diffusion with resetting process using the law of total expectation. In Section~\ref{sec:ergodicity} we rigorously study the ergodic properties of the mean and the TAMSD. Finally, we summarize our findings and discuss potential avenues for future research in Section~\ref{sec:conclusion}. An appendix is added to provide supporting derivations.

\section{Diffusion with resetting}
\label{sec:diffusiong-with-resetting}

\subsection{Explicit solution to diffusion with resetting}

The diffusion of a particle governed by a drift and which is subjected to resetting is described by the following Langevin equation
\bea
d x(t) &=(1-Z_{t}) \left( \mu dt+ \sigma dW \right)+Z_{t} \left( x_0-x(t) \right),
\label{eq:diffusion-restart-microscopic}
\eea
where $x(t)$ is the position of the particle at time $t$, $dt$ denotes the infinitesimal time increment and
$dW$ is an an infinitesimal Wiener increment, which is a normal variate with $\langle dW_t \rangle=0$ and $\langle dW_t dW_s \rangle =\delta(t-s)dt$. Here, $\mu$ and $\sigma$ are called the drift and noise amplitude. Resetting is introduced with a random variable $Z_{t}$ which takes the value $1$ when there is a resetting event in the time interval between $t$ and $t +dt$; otherwise, it is zero. Without any loss of generality, we also assume that resetting brings the particle back to its initial condition which is the origin, i.e, $x(0)=x_0=0$.

The solution to \eqref{eq:diffusion-restart-microscopic} can be found by interpreting the diffusion as a renewal process: each resetting event renews the process at $x_0$ and between two such consecutive renewal events, the particle undergoes the simple diffusion. Thus, between time points $0$ and $t$, only the last resetting event, occurring at the point 
\begin{align}
    t_{l}(t) = \max_{k \in \left[0,t\right]} k: \{ Z_{k} = 1 \},
    \label{eq:diffusion-restart-solution-1-0}
\end{align}
is relevant and the solution to  \eqref{eq:diffusion-restart-microscopic} reads
\begin{align}
      x(t) &= \mu \bigg(t - t_l(t)\bigg) + \sigma \bigg(W(t) - W(t_l(t))\bigg).
      \label{eq:diffusion-restart-solution-2-0}
\end{align}
Writing stochastic solutions on a single trajectory level in the presence of resetting is quite useful as it allows to directly apply the \textit{law of total expectation} to the random variable such as the position $x(t)$ and study the statistical properties of the process such as its moments and/or  other quantiles, given the distribution of $t_l(t)$.

\subsection{Exponential reset time distribution/Poissonian resetting}
In what follows we will assume stochastic resetting so that the probability for a reset event is given by $P(Z_{t} = 1) = rdt$. In the limit when $dt \to 0$, this corresponds to an exponential resetting time density $f_r(t)=re^{-rt}$, and $t_l$ is distributed according to
\begin{align}
f(t_{l}|t)=\delta(t_{l}) e^{-rt}+re^{-r(t-t_{l})},
\label{last-time-pdf}
\end{align}
such that $\int_0^t\,dt_l f(t_l|t)=1$.

\begin{figure*}[h]

\includegraphics[width=14cm]{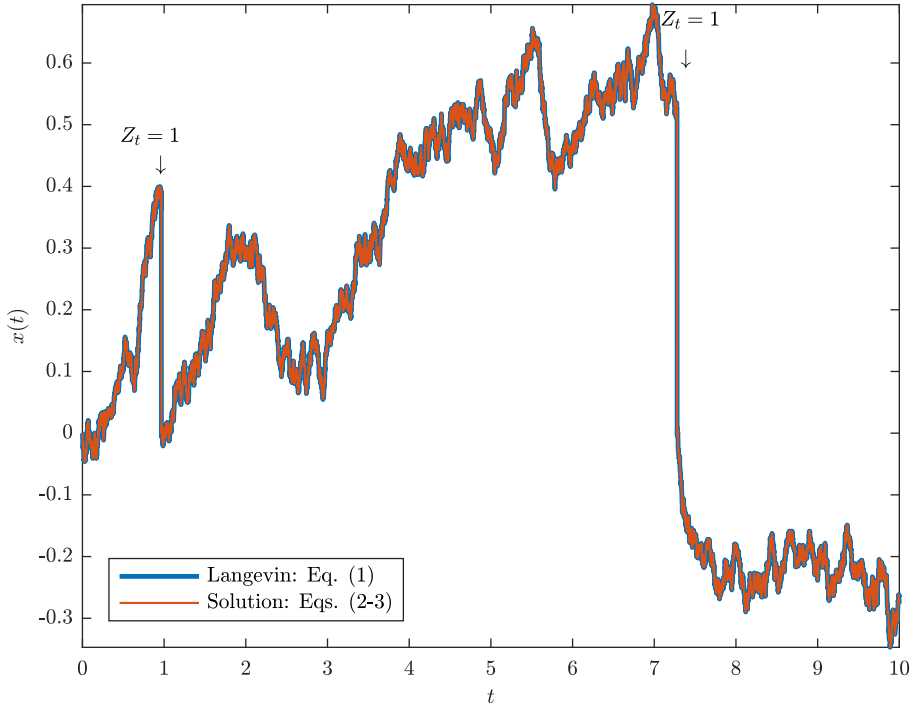} 
\caption{\textbf{Simple depiction of a diffusing trajectory with resetting}. The particle diffuses until a reset event characterized by $Z_t = 1$ occurs. At this moment, position of the particle is reset to $x_0=0$ and the dynamics is renewed. The blue line describes a numerical simulation of the Langevin equation~\eqref{eq:diffusion-restart-microscopic}, whereas the orange line is the analytical solutions~\eqref{eq:diffusion-restart-solution-1-0} and~\eqref{eq:diffusion-restart-solution-2-0} taken together. In this example, we set $\mu = 0.05$, $\sigma^2 = 0.02$ and   \label{fig:solution}}
\end{figure*}

Intuitively, the first term on the RHS corresponds to the scenario when there is no resetting event up to time $t$ while the second one accounts for multiple resetting events. 
Nonetheless, we stress out that the solution also holds for complex restart time distributions with a straightforward generalization of \eqref{last-time-pdf} that can be obtained from Refs. \cite{Pal-time-dep,Chechkin}. Fig.~\ref{fig:solution} compares the solution~\eqref{eq:diffusion-restart-solution-1-0} and~\eqref{eq:diffusion-restart-solution-2-0} with the Langevin simulation and shows a perfect match.

As will be seen below, in order to apply the law of total expectation to the displacement of a diffusing particle subject to stochastic resetting, we will require the following statistical properties of the reset time distribution: the first moment $\langle t_l(t) \rangle$, the variance $\mathrm{var}\left[t_l(t) \right]$,  the autocovariance $\mathrm{cov}\left[t_l(t),t_l(t+\Delta) \right]$ between the time of the last resetting event belonging to two different periods $t+\Delta$ and $t$ and the average of the minimum between $t_l(t+\Delta)$ and $t$, i.e. $\langle \min \left\{t_l(t+\Delta),t \right\} \rangle$.

\subsubsection{Moments.---}
In general, the $m$-th moment of $t_l(t)$ can be found as $\langle t_l^m(t) = \int_0^t\,dt_l~ t_l^m f(t_l|t)=t^m e^{-r t} (-r t)^{-m} (\Gamma (m+1,-r t)-m \Gamma (m))$. In particular, the first moment reads 
\begin{align}
     \langle t_l(t) \rangle &= t-\frac{1-e^{-rt}}{r},
     \label{eq:first-moment-t_l}
     \end{align}
whereas the variance is given by
\begin{align}
   \mathrm{var}\left[t_l(t) \right] &= \langle t_l^2(t) \rangle - \langle t_l(t) \rangle^2 \nonumber\\
&= \frac{1- 2rt e^{-rt} - e^{-2rt}}{r^2}.
  \label{eq:second-moment-t_l}
\end{align}

\subsubsection{Autocovariance.---}
\begin{figure}
    \centering
    \includegraphics[width=0.7\textwidth]{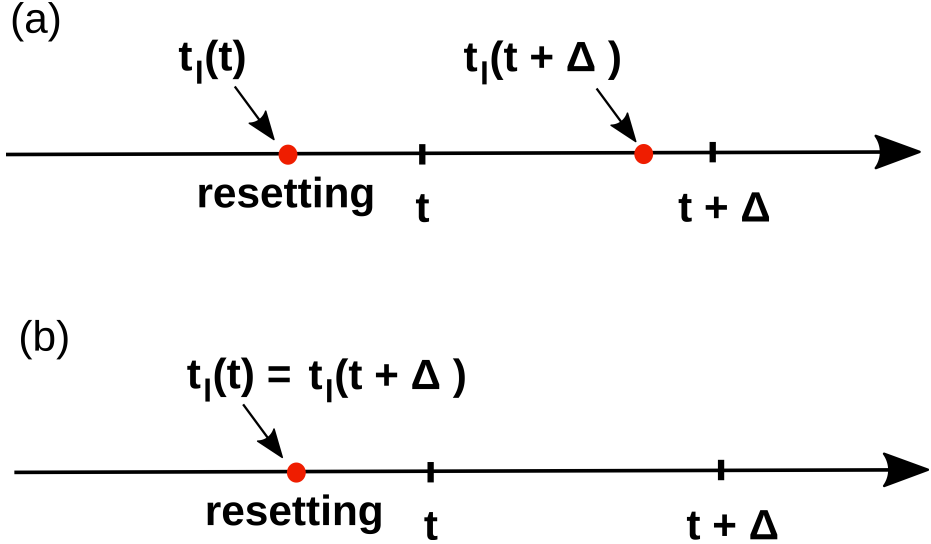}
    \caption{\textbf{Computation of autocovariance for resetting events.} We depict the two possible scenarios while computing $\mathrm{cov}\left[t_l(t),t_l(t+\Delta) \right]$. Red circles indicate last resetting events given an observation time. Panel (a) depicts a scenario when there is atleast one resetting event between the two given time scales $t$ and $t+\Delta$, and panel (b) considers the complementary situation i.e., when there is none.} 
    \label{covariance-tl}
\end{figure}

We now present the derivation of the autocovariance given by $\mathrm{cov}\left[t_l(t),t_l(t+\Delta) \right]$ which is a little bit more subtle as it relies on the properties of the process $t_l(t+\Delta) t_l(t)$.  First note from the law of total expectation that 
\begin{align}
    \left\langle t_l(t+\Delta) t_l(t) \right\rangle &=     \left\langle \left\langle t_l(t+\Delta) t_l(t)|t_l(t) \right\rangle_{t_l(t+\Delta)} \right\rangle_{t_l(t)},
\end{align}
where $\left\langle t_l(t+\Delta)\,t_l(t)|t_l(t) \right\rangle_{t_l(t+\Delta)}$ is the conditional average of $t_l(t+\Delta)\,t_l(t)$ provided that we know $t_l(t)$. This observable can be computed by observing that there are two possibilities (see Fig. \ref{covariance-tl}): (a) a last resetting event occurs at time $t_l(t+\Delta)$ between $t$ and $t+\Delta$ and (b) there is no resetting between $t$ and $t+\Delta$  such that $t_l(t+\Delta)=t_l(t)$. Combining the possibility of these two events, one can write
\begin{align}
    t_l(t+\Delta) t_l(t) | t_l(t) &= \begin{cases}
  t_l^2(t), & \textrm{if no reset between $t$ and $t+\Delta$,} \\
   t_l(t)  t_1, & \textrm{otherwise,} 
\end{cases}
\end{align}
where we have denoted the last resetting time point between $t$ and $t+\Delta$ as $t_1$.  
The probability of no reset between $t$ and $t+\Delta$ is $e^{-r\Delta}$ [see Fig. \ref{covariance-tl}b], whereas the probability that the last reset occurred at a point $t_1$ (with $t < t_1 \leq t+\Delta$) is $r e^{-r(t-t_1)}$ [see Fig. \ref{covariance-tl}a]. Therefore, the conditional average can be found as 
\begin{align}
    \left\langle t_l(t+\Delta) t_l(t)|t_l(t) \right\rangle_{t_l(t+\Delta)} &= t_l^2(t) e^{-r\Delta} + t_l(t) \int_t^{t+\Delta} t_1 r e^{-r(t+\Delta-t_1)} d t_1 \nonumber \\
    &= t_l^2(t) e^{-r\Delta} + t_l(t) \frac{1}{r} \left[  e^{-r\Delta} (1-rt) + r(t+\Delta) - 1\right]. 
\end{align}

Averaging the above equation with respect to $t_l(t)$ we get that
\begin{align}
    \left\langle t_l(t+\Delta) t_l(t) \right\rangle &=  \langle t_l^2(t) \rangle e^{-r\Delta} + \langle t_l(t) \rangle \frac{1}{r} \left(  e^{-r\Delta} (1-rt) + r(t+\Delta) - 1\right)\nonumber \\
    &= t(t+\Delta) - \frac{2t+\Delta}{r} + \frac{1}{r}\left( e^{-rt}(t+\Delta) - e^{-r(t+\Delta)}t \right) \nonumber\\& + \left( \frac{1-e^{-rt}}{r^2}\right)\left( 1 + e^{-r\Delta} \right),
\end{align}
which implies that
\begin{align}
\mathrm{cov}\left[t_l(t),t_l(t+\Delta) \right] &= \langle t_l(t),t_l(t+\Delta) \rangle - \langle t_l(t) \rangle \langle t_l(t+\Delta)\rangle\nonumber  \\
&= e^{-r\Delta}\frac{1-2te^{-rt}-e^{-2rt}}{r^2}\nonumber \\
&= e^{-r\Delta} \,   \mathrm{var}\left[t_l(t) \right].
    \label{eq:cross-moment-t_l}
\end{align}

\subsubsection{Computation of $\langle \min \left\{t_l(t+\Delta),t \right\} \rangle$.---}
We now turn our attention to compute $ \langle \min \left\{t_l(t+\Delta),t\right\} \rangle$. Generically, this can be presented in terms of the cumulative distribution function the reset time, i.e., $F(y|t) = \int_0^t\,dt_l f(t_{l}|t)$. Since $t_l(t+\Delta)$ is a non-negative random variable, the minimum between this random variable and a constant $t$ can be written as
\begin{align}
  \langle \min\{t_l(t+\Delta),t\} \rangle &= \int_0^t\,dy\,Pr \left[t_l(t+\Delta) \geq y\right]\nonumber\\
  &=\int_0^t\,dy\,\left[1-Pr \left(t_l(t+\Delta) < y\right)\right]\nonumber \\
   &=\int_0^t(1 - F(y|t+\Delta)) dy\nonumber \\
    &= \int_0^t dy - \int_0^s F(y|t+\Delta)\,dy\nonumber \\
   &= t - \int_0^t F(y|t+\Delta)\, dy.
\end{align}
In the case of Poissonian resetting, we have $F(y|t+\Delta) = e^{-r(t+\Delta-y)}$, and thus
\begin{align}
    \langle \min \left\{t_l(t+\Delta),t \right\} \rangle &= t-e^{-r\Delta} ~\frac{ 1 - e^{-rt}}{r}.
    \label{eq:min-t_l}
\end{align}

 \section{The law of total expectation in diffusion with resetting}
 \label{sec:law-of-total-expectation}
 
\subsection{Moments of diffusion with resetting}
Using the law of total expectation, the $m$-th moment of diffusion with stochastic resetting can be obtained by raising \eqref{eq:diffusion-restart-solution-2-0} to the $m$-th power, then averaging with respect to the Wiener noise and $f(t_l|t)$ respectively. This can formally be written as
\begin{align}
 \langle x^m(t) \rangle &=   \bigg\langle \left\langle x^m(t) |t_l \right\rangle_{W} \bigg\rangle_{t_l},
\end{align}
where $\left\langle x^m(t) |t_l \right\rangle_{W}$ is the conditional average of $x^m(t)$ provided that we know $t_l(t)$ followed by $\langle \cdot \rangle_{t_l}$ that denotes an averaging with respect to the random variable $t_l$. The first conditional average is with respect to the Wiener noise and requires the following ingredients: $\langle W(t) \rangle=0$,$~\mathrm{Var}\left[W(t)\right]=t$,~
$\langle W(t) W(s)\rangle=\min\left\{s,t\right\}$. 
The second averaging follows directly from the reset time distribution $f(t_l|t)$ given by (\ref{last-time-pdf}).

Using this procedure the first moment $\left\langle x(t)\right\rangle$ can be computed in the following way
\begin{align}
    \langle x(t) \rangle &=  \mu \bigg(t - \langle t_l(t) \rangle \bigg) = \mu \frac{1-e^{-rt}}{r},
    \label{eq:bm-reset-first-moment}
\end{align}
where we have used Eqs.  (\ref{eq:diffusion-restart-solution-2-0}) and (\ref{eq:first-moment-t_l}). 
Similarly, the second moment is obtained by squaring (\ref{eq:diffusion-restart-solution-2-0}) and then systematically taking the averages
\begin{align}
  \langle x^2(t) \rangle &= \mu^2 \bigg( t^2 - 2 \langle t_l(t) \rangle t + \langle t_l^2(t) \rangle\bigg) + \sigma^2 \bigg( t - \langle t_l(t) \rangle  \bigg) \nonumber\\
  &= \frac{2\mu^2}{r} \left(\frac{1-e^{-rt}}{r} - t e^{-rt}\right) + \sigma^2 \left( \frac{1- e^{-rt}}{r}\right).
\label{eq:bm-reset-second-moment}
\end{align}
This result is the same as the one obtained from the correpsponding Fokker-Planck equation for a Brownian motion with constant drift, see Eq.~(31) from Ref.~\cite{math2021} ($\mu\rightarrow V$, $\sigma^2\rightarrow2\mathcal{D}$, $x_0=0$), as it should be. 
Finally, the variance is given by
\begin{align}
 \mathrm{var}\left[x(t)\right] =\langle x(t)^2\rangle - \langle x(t) \rangle^2 &= \mu^2  \mathrm{var}\left[t_l(t)\right] + \sigma^2 \bigg( t - \langle t_l(t) \rangle  \bigg).
 \label{eq:bm_restart_variance}
\end{align}

\subsection{Autocorrelations in diffusion with stochastic resetting}

In this section, we analyze the autocorrelation function $\mathrm{corr}\left(x(t), x(t+\Delta)\right)$ which is formally defined as
\begin{align}
    \mathrm{corr}\left[x(t), x(t+\Delta)\right] &= \frac{\mathrm{cov}[x(t), x(t+\Delta)]}{\sqrt{\mathrm{var}[x(t)]} \sqrt{\mathrm{var}[x(t+\Delta)]}},
    \label{eq:autocorrelation}
\end{align}
where recall that the autocovariance is given by
\bea
\mathrm{cov}[x(t), x(t+\Delta)] &\equiv \langle x(t) x(t+\Delta) \rangle -\langle  x(t+\Delta) \rangle \langle x(t)  \rangle .
\label{eq:bm_restart_covariance}
\eea
In words, the autocorrelation function is simply the Pearson correlation between two different time points, $t$ and $t+\Delta$ of a given observable. The autocorrelation coefficient is bounded between $-1$ and $1$, with $\mathrm{corr}[x(t), x(t+\Delta)]=1$ meaning perfect correlation. That is, a change in the present position of a trajectory is exactly proportional to the change in the future position. When $\mathrm{corr}[x(t), x(t+\Delta)]=-1$ then increments in $x(t)$ imply decrements of the same magnitude in $x(t+\Delta)$. Lastly, when $\mathrm{corr}[x(t), x(t+\Delta)]=0$, then trajectories are uncorrelated and the current position of the particle does not offer any information about the future positions.

The solution~\eqref{eq:diffusion-restart-solution-1-0} can be coupled with the law of total expectation to calculate the autocovariance $\mathrm{cov}\left[x(t), x(t+\Delta)\right]$ and the autocorrelation as we show below. Similar to the analysis made for the moments, the law of total expectation implies that the non-normalized autocorrelation, i.e., the average of $C_\Delta (t) \equiv x(t) x(t+\Delta)$, can be written as
\begin{align}
 \langle C_\Delta (t) \rangle &=   \left\langle \left\langle x(t) x(t+\Delta) |t_l \right\rangle_{W} \right\rangle_{t_l},
\end{align}
where the first averaging is with respect to the Wiener noise i.e.,
\begin{align}
 \left\langle x(t) x(t+\Delta) |t_l \right\rangle_{W} &= \mu^2 \bigg( t - t_l(t) \bigg) \bigg( t+ \Delta - t_l(t+\Delta) \bigg) + \sigma^2 \bigg(t - \min\left\{t_l(t+\Delta),t \right\}\bigg).
\end{align}
Next, we average the above equation with respect to $t_l$ to derive a general form for the average of the function to find
\begin{align}
  \langle C_\Delta (t) \rangle 
    & = \mu^2 \left[ t(t+\Delta)-t \langle t_l(t+\Delta) \rangle-(t+\Delta) \langle t_l(t) \rangle+\langle t_l(t+\Delta)t_l(t) \rangle \right]\nonumber\\& + \sigma^2 \left[ t-\langle \text{min}(t_l(t+\Delta),t) \rangle \right],
    \label{eq:non-normalized-correlation-function}
\end{align}
and is independent of the specific form of $f(t_l|t)$. For Poissonian resetting, this reads
\begin{align}
 \langle C_\Delta (t) \rangle =
\frac{\mu^2}{r^2} \bigg[ 1+e^{-r\Delta} -e^{-rt}-e^{-r(t+\Delta)}(1+2rt) \bigg]+\frac{\sigma^2}{r} e^{-r \Delta} (1-e^{-rt}),
\end{align}
where we have used \cref{eq:cross-moment-t_l,eq:min-t_l}. 
Moreover, the autocovariance function can be obtained using (\ref{eq:bm_restart_covariance})
\begin{align}
\mathrm{cov}[x(t), x(t+\Delta)] 
&= \mu^2  \mathrm{cov}[t_l(t), t_l(t+\Delta)] + \sigma^2 \left[ t-\langle \text{min}(t_l(t+\Delta),t) \rangle \right].
\label{eq:resetting-autocovariance}
\end{align}
For Poissonian resetting, this simplifies to
\begin{align}
\mathrm{cov}[x(t), x(t+\Delta)] &= e^{-r\Delta}\bigg[\frac{\mu^2}{r^2}\left(1-2rte^{-rt}-e^{-2rt}\right) +\frac{\sigma^2}{r}\left( 1- e^{-rt} \right) \bigg].
\label{eq:bm-reset-ACV}
\end{align}
In the case of no-resetting, it is known that the autocovariance diverges because the underlying drift-diffusion process is non-stationary. However, the non-equilibrium steady state induced by Poissonian resetting (exact derivation is shown in the next section) also leads to a convergent autocovariance. We remark that
one can further insert \eqref{eq:resetting-autocovariance} and \eqref{eq:bm_restart_variance}  in~\eqref{eq:autocorrelation} to obtain a general expression for the autocorrelation function under general reset time distribution. 

For the Poissonian resetting case, it can be shown that the steady-state autocorrelation is given by
\begin{align}
  \text{corr}\left[\Delta \right]\equiv \lim_{t \rightarrow \infty} \text{corr}\left[x(t), x(t+\Delta)\right]  & = e^{-r\Delta}.
  \label{eq:normalized-correlation}
\end{align}
This implies that for any finite $\Delta$ there will be a positive correlation between present and future positions, however as $\Delta \to \infty$ the trajectories become uncorrelated. A similar expression was derived in \cite{oshanin} for simple diffusion without any drift. There the autocovariance function under Poissonian resetting was studied in terms of the autocorrelation of the underlying non-reset process. It was argued that the observed convergence is a result of the disjoint ``renewal'' intervals which cause independence between trajectories belonging to different intervals. The same reasoning also applies here for the autocorrelation function of the drift-diffusion process under stochastic resetting.

\section{Ergodic properties of diffusion with stochastic resetting}
\label{sec:ergodicity}

\subsection{Preliminaries}
\label{mixing-prelim}

In this section, we study the ergodic properties of the drift-diffusion process with stochastic resetting in detail. We start by discussing the \textit{mixing property} of the process. In simple words, mixing implies that the fraction of the time spent by a particle within a particular region in the \textit{phase space} is proportional to the volume of that region. This entails long time independence of the behavior of single trajectories on their previous values. Every process that is mixing also has ergodic and stationary observables, but the opposite is not necessarily true~\cite{petersen1989}. Therefore, mixing is usually known as a much stricter concept than ergodicity~\cite{shalizi2006advanced}. In this section, we first unravel the mixing properties of diffusion with resetting, and then show how it helps us to build intuitions on the ergodic behavior of the said system.

Mathematically, mixing can be analyzed in various ways. One simple procedure for proving that ``\textit{a process is mixing}'' is by studying the decay of autocorrelations, once we have shown that the process is stationary. That is, a stationary process is mixing if and only if its autocorrelation function decays to zero, i.e., $\lim_{\Delta_\to\infty}\text{corr}\left[\Delta \right]=0$~\cite{liu2017correlation,maruyama1970infinitely}.
The reasoning behind this procedure is that, once the system is in a stationary state, the autocorrelation function quantifies the decay in the dependence of the particle's position from its previous positions in time. 

In our system of diffusion with resetting, note that the autocorrelation converges to zero i.e., $\lim_{\Delta_\to\infty}\text{corr}\left[\Delta \right]=0$ [from \eqref{eq:normalized-correlation}]. Thus what remains to be shown is that the process is stationary. Stationarity of simple diffusion with stochastic resetting is a well known fact and its various features have been investigated in various studies. With added drift, a similar stationarity can also be rendered. For completeness, we present the derivation here. We use the standard renewal formalism that allows one to write the propagator of the reset-process in terms of the reset-free process. Setting $P_{r}(x,t)$ as the probability density function (PDF) that governs the particle position $x$ in time $t$, the renewal equation reads
\begin{align}\label{renewal eq}
    P_{r}(x,t)=e^{-rt}P_{0}(x,t)+\int_{0}^{t}re^{-rt'}P_{0}(x,t')\,dt,
\end{align}
where 
\begin{align}\label{shifted gaussian}
    P_{0}(x,t)=\frac{1}{\sqrt{2\pi \sigma^2 t}}e^{-\frac{(x-\mu t)^2}{2\sigma^2 t}},
\end{align}
is the PDF of the drift-diffusion equation without resetting ($r=0$) and the initial position is $x_0=0$ \cite{math2021}), i.e., $P_{r}(x,t=0)=\delta(x)$. The stationary PDF, $P_r^{*}(x)=\lim_{t\rightarrow\infty}P_{r}(x,t)$ then takes the following form
\begin{align}\label{ness}
    P_r^{*}(x) &=\frac{r}{\sqrt{\mu^2+2\sigma^2r}}\exp\left(\frac{\mu}{\sigma^2}x-\frac{\sqrt{\mu^2+2\sigma^2 r}}{\sigma^2}|x|\right),
\end{align}
which is a time-independent function. This clearly states drift-diffusion with stochastic resetting is mixing. This result is unsurprising since each reset renews the process. As a result, positions of the particle belonging to two different renewal intervals are independent. We provide an alternate procedure for showing the mixing properties by using the spectral features of the process in \ref{app:relaxation}.

As stated before, mixing is strongly related to the concept of ergodicity. An observable $f(x(t))$, resulting from a continuous time stochastic process, is ergodic if its time average,
\begin{align}
     \overline{f} =  \lim_{T\to\infty}\frac{1}{T}\int_0^T f(x(t)) dt,
\end{align}
where $T$ is some observation time, is equal to the ensemble average 
\begin{align}
    \langle f \rangle = \int_{\Omega} f(x) P(x) dx,
\end{align}
where $\Omega$ is the sample space and $P(x)$ is the stationary PDF of the process. The ergodic hypothesis states that an observable $f(x)$ is ergodic if the first moment of $ \overline{f}$ is equal to the ensemble average of the process, i.e.,
\begin{align}
 \langle  \overline{f} \rangle = \langle f \rangle,
\end{align}
and that its variance $\mathrm{Var}\left[ \overline{f}\right]$ is zero. Here, averaging over noise realizations is denoted by the angular brackets,
while time averaging is shown by the overline.

By definition, mixing implies ergodicity of the ensemble average, the mean squared displacement and the autocorrelation of the same process. In what follows, we will utilize this approach to show that the mean of the initial process of diffusion under stochastic resetting is ergodic. Afterwards, we will argue that the same intuition follows for the ergodic behavior of the second moment and the (non-normalized) autocorrelation. Subsequently, we will apply the results to investigate the ergodic properties of another observable namely the time-averaged mean-squared displacement (TAMSD), a quantity widely used in single particle tracking experiments.  

\subsection{Ergodicity of sample mean under stochastic resetting}
\begin{figure*}[h]
\includegraphics[width=18cm]{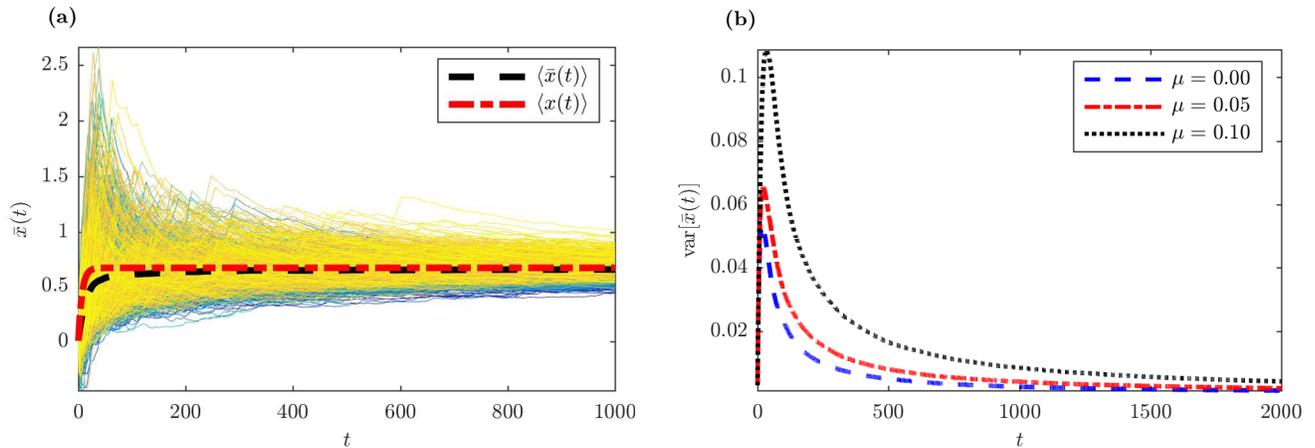} 
\caption{\textbf{Ergodicity of the \textit{sample mean} in diffusion with stochastic resetting}. \textbf{(a)} The time-average of the mean $ \overline{x}(t)$ for $N=1$  as a function of time for $10^3$ realizations (thin lines). The black dashed line is the average of $ \overline{x}(t)$ across the realizations, whereas the red dash-dotted line is the analytical value for the ensemble average $\langle x(t) \rangle$. In the simulation we set $\mu = 0.1$. \textbf{(b)} Variance $\mathrm{var}\left[ \overline{x}(t) \right]$ of the time-averaged mean as a function of time for various drift rates. Asymptotic convergence of the variance to zero is observed. Parameters in \textbf{(a-b)}: We set $r = 0.15$, $\sigma^2 = 0.03$, $\Delta = 10$, $dt = 0.1$ and average the results across $10^3$ realizations.  \label{fig:mean-ergodicity}}
\end{figure*}

The sample mean $\langle x(t) \rangle_N$ over a finite sample of $N$ trajectories is simply the sum of their positions divided by the number of trajectories, i.e.
\begin{align}
    \langle x(t) \rangle_N = \frac{\sum_i^N x_i(t)}{N}.
\end{align}
Notice that as $N\to\infty$, the sample mean is basically the ensemble average. Therefore, in empirical studies, one always uses the mean as an approximation for the ensemble average. However, in a non-ergodic system, unless the sample is very large, there are significant differences between the mean and the ensemble average. This phenomenon is particularly relevant in economics studies, where one usually uses the sample mean as an approach to reduce the potential net-negative impact of fluctuations on asset prices~\cite{peters2013ergodicity,stojkoski2019cooperation,stojkoski2021evolution}.
The time average of the mean is defined as
\begin{align}
     \overline{x}_N &= \lim_{T\to\infty} \frac{1}{T}\int_0^T  \langle x(t) \rangle_N ~ dt,
    \label{eq:time-average-sample-mean}
\end{align}
with the ensemble average given by (for any $N$)
\begin{align}
    \langle  \overline{x}_N \rangle &= \lim_{T\to\infty} \frac{1}{T}\int_0^T  \langle \langle x(t) \rangle_N \rangle ~ dt.
\end{align}
 Each individual trajectory is independent and identically distributed. This indicates that if the special $N=1$ case is ergodic, then the same property will hold for any arbitrary $N$. Therefore, let us analyze the $N=1$ case. In this context, the average of $ \overline{x}$ (where we omit the $N$ notation simply because we are analyzing a single trajectory) can be found as 
\begin{align}
    \langle  \overline{x} \rangle &= \lim_{T\to\infty} \frac{1}{T}\int_0^T \langle x(t) \rangle dt \nonumber\\& = \frac{\mu}{r}\lim_{T\to\infty} \frac{1}{T}\int_0^T  \left(1-e^{-rt}\right) dt = \frac{\mu}{r}.
    \label{eq:time-average-mean}
\end{align}
 On the other hand, the stationary ensemble average of the same observable can be found using \eqref{eq:bm-reset-first-moment} as 
 \begin{align}
  \langle x \rangle =  \lim_{t\to\infty} \langle x(t) \rangle = \frac{\mu}{r}
 \end{align}
Hence, we have 
\begin{align}
 \langle  \overline{x} \rangle = \langle x \rangle   ,
\end{align}
for any $\mu $. In Fig.~\ref{fig:mean-ergodicity}(a) we numerically illustrate  this finding by fixing $\mu$,~$\sigma$ and $r$ and plotting $ \overline{x}(t) = \overline{x}_{N=1}(t)= \frac{1}{T}\int_0^T  x(t) dt$  as a function of time for various realizations of diffusion with stochastic resetting. The figure also shows the mean time-average $\langle  \overline{x}(t) \rangle$, averaged across $10^3$ realizations (dashed black line). The mean time average evolves over time, and eventually converges to the stationary ensemble average $\mu/r$ (shown by dash-dotted red line), thus conforming to our result.

Next, let us examine the variance $\mathrm{var}\left[ \overline{x}_{N=1}\right]$ of the time average. This quantity is given by
\begin{align}
 \mathrm{var}\left[ \overline{x}_{N=1}\right] &= \lim_{T\to\infty} \frac{1}{T^2}\int_0^T d t_2 \int_0^T d t_1 \mathrm{cov}\left[ x(t_1),x(t_2) \right],
\end{align}
where we estimate the autocovariance function $\mathrm{cov}\left[ x(t_1),x(t_2) \right]$ from
\eqref{eq:bm_restart_covariance}. The expression in the above equation will always converge to $0$ because the integral grows proportionally to $T$. We visualize this phenomenon numerically in Fig.~\ref{fig:mean-ergodicity}(b) by displaying the variance of the time averaged mean $ \mathrm{var}\left[ \overline{x}(t)\right]$ as a function of time for various drift rates. In each case, the variance converges to zero. Altogether, this implies that the \textit{sample mean is an ergodic observable} in \textit{drift-diffusion with stochastic resetting}. However, the mixing nature of the initial process of diffusion with stochastic implies that observables such as the TAMSD will be non-ergodic. This will be the focus of our next section.

\subsection{Ergodicity of  TAMSD under stochastic resetting}

\subsubsection{Definition and results:~}

The TAMSD of a single particle whose dynamics is governed by a stochastic process $x(t)$ is defined as \cite{TAMSD-review-1,TAMSD-review-2,TAMSD-review-3}
\begin{align}
   \delta^2(\Delta,T) &= \frac{1}{T-\Delta}\int_0^{T-\Delta} \left[ x(t+\Delta) - x(t) \right]^2 dt,
\end{align}
where $\Delta$ is the lag time and $T$ is the length of the time-series as before.
The TAMSD is often used to measure the portion of the system explored by a particle over time.  As such it is widely applicable in tracking particle dynamics in various domains and its analytical features have been extensively investigated over the last years for a variety of stochastic processes~\cite{TAMSD-review-1,TAMSD-review-2,TAMSD-review-3,suzuki,he_prl,sokolov_prl,akimoto_prl}.

Ergodicity necessitates equality between the second moment (mean-squared displacement) and the long time average TAMSD, $\delta^2(\Delta)$, for a fixed $\Delta$, i.e.,
\begin{align}
\delta^2(\Delta)=\lim_{T\rightarrow\infty}\delta^2(\Delta,T) =\langle x^2(\Delta) \rangle.  
\label{TAMSD-ergodicity-equivalence}
\end{align}
In the case of free Brownian motion for long measurement time, both ensemble and time averaged MSD are the same. That is, the diffusion coefficient obtained from an individual trajectory is the same as the one obtained from an ensemble of particles during a period $\Delta$, and therefore the TAMSD in simple Brownian motion is ergodic. 

We hereby point out that the long time TAMSD in diffusion with resetting can be rewritten as
\begin{align}
    \delta^2(\Delta) &= \lim_{T \to \infty} \int_0^{T-\Delta} \left[   x^2(t+\Delta) + x^2(t) -  2 x(t+\Delta) x(t) \right] dt\nonumber \\
    &= 2 \left(  \overline{x^2} -  \overline{C}_\Delta \right), \label{TAMSD-reset}
\end{align}
where 
\begin{align}
 \overline{x^2} & \equiv \lim_{T\to\infty} \frac{1}{T}\int_0^T x^2(t)\, dt \nonumber\\& = 
 \frac{2\mu^2}{r^2} + \frac{\sigma^2}{r},
    \label{eq:time-average-second-moment}
\end{align}
is the time-average of the squared displacement (corresponds to the choice of $f(x(t)) = x^2(t)$) and
\begin{align}
  \overline{C}_\Delta & \equiv \lim_{T\to\infty} \frac{1}{T}\int_0^T x(t) x(t+\Delta)\, dt \nonumber\\
    &= \frac{\mu^2}{r^2} + e^{-r\Delta}\left( \frac{\mu^2}{r^2} + \frac{\sigma^2}{r}\right), 
    \label{eq:time-average-autocorrelation}
\end{align}
is the time-average of the autocorrelation for the choice $f(x(t)) = x(t)x(t+\Delta)$. Note that both the derivations use the stochastic solution $x(t)$ for drift-diffusion with stochastic resetting namely \eqref{eq:diffusion-restart-solution-2-0}.
This is a neat formulation as it allows us to directly quantify the long time TAMSD through the time-averages of the squared particle position and the autocorrelation.

\begin{figure*}[t!]

\includegraphics[width=18cm]{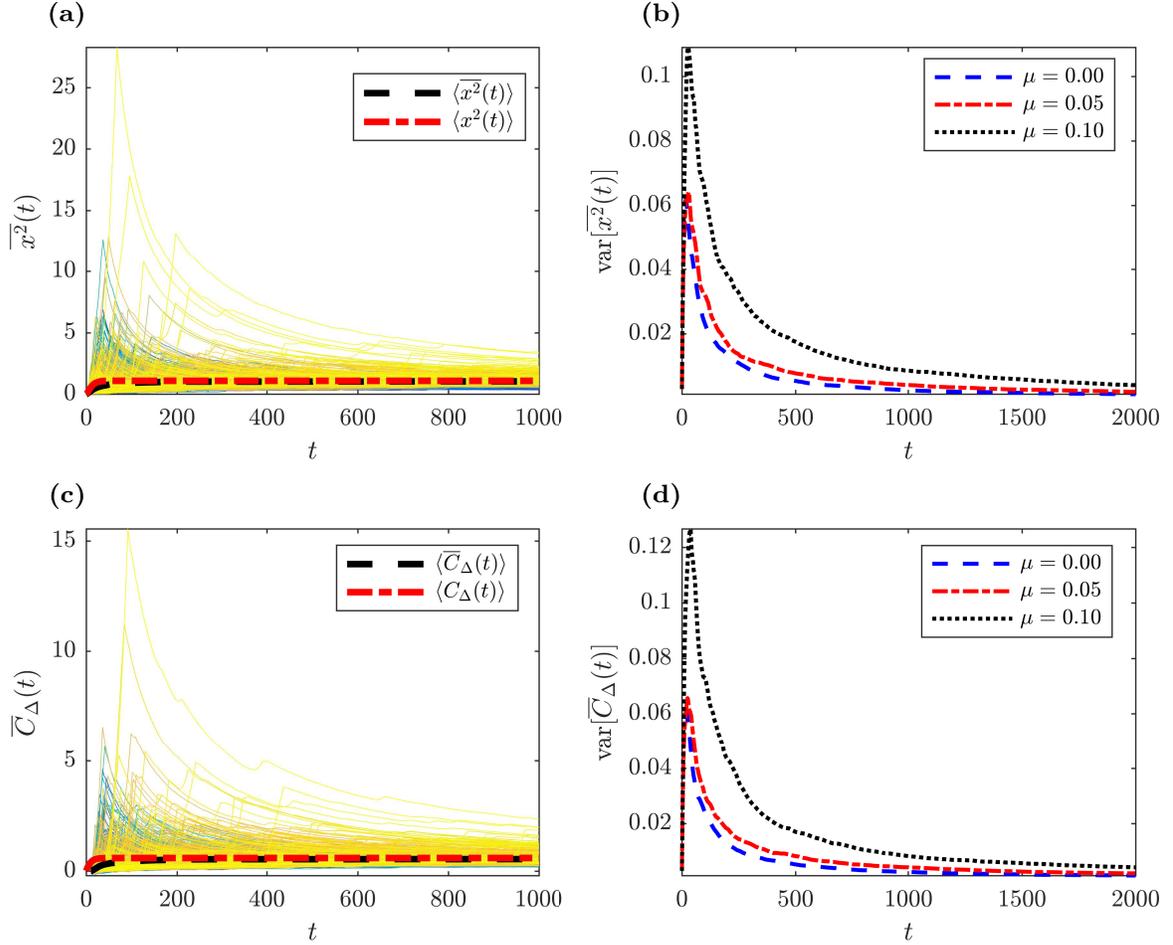} \caption{\textbf{Ergodicity of the \textit{second moment} and \textit{autocorrelation function} in diffusion with stochastic resetting}. \textbf{(a)} The time-average of the second moment $ \overline{x^2}(t)$ as a function of time for $10^3$ realizations (thin lines). The black dashed line is the average of $ \overline{x^2}(t)$ across the realizations, whereas the red dash-dotted line is the analytical value for the ensemble average $\langle x^2 \rangle$. In the simulation we set $\mu = 0.1$.
\textbf{(b)} Variance $\mathrm{var}\left[ \overline{x^2}(t) \right]$ of the time-averaged second moment as a function of time for various drift rates.
 \textbf{(c)} Same as \textbf{(a)} but for the time-average of the non-normalized autocorrelation $\overline{C}_\Delta (t)$. \textbf{(d)} Same as \textbf{(b)} but for the time-average of $\overline{C}_\Delta (t)$.  Parameters for
 \textbf{(a-d)}: We set $r = 0.15$, $\sigma^2 = 0.03$, $\Delta = 10$, $dt = 0.1$ and average the results across $10^3$ realizations.  \label{fig:autocorrelation-ergodicity}}
\end{figure*}

The results in Eqs.~\eqref{eq:time-average-second-moment} and~\eqref{eq:time-average-autocorrelation} imply that these two observables are ergodic
\begin{align}
      \overline{x^2} &= \lim_{t\to\infty}  \langle x^2(t) \rangle,  \\
       \overline{C}_\Delta &= \lim_{t\to\infty} \langle x(t)x(t+\Delta) \rangle ,
\end{align}
where the correspondence has been made by recalling the expressions for the square moment from \eqref{eq:bm-reset-second-moment} and autocorrelation from \eqref{eq:bm-reset-ACV} respectively. 
Indeed, this follows from the previous finding that the drift-diffusion process is mixing. In order to confirm our prediction, we visualize the temporal evolution 
of the observables $ \overline{x^2}(t)$ and $  \overline{C}_\Delta(t)$ for a fixed choice of parameters. See Fig~\ref{fig:autocorrelation-ergodicity}(a) and Fig~\ref{fig:autocorrelation-ergodicity}(c). Moreover, in Fig~\ref{fig:autocorrelation-ergodicity}(b) and Fig~\ref{fig:autocorrelation-ergodicity}(d) we show the temporal decay of the variance for the same observables and their convergence to zero i.e., $\lim_{t\to\infty}\mathrm{var}\left[ \overline{x^2}(t)\right] = 0$ and $\lim_{t\to\infty}\mathrm{var}\left[   \overline{C}_\Delta (t)\right]=0$ as it should be in the case of ergodicity.

\begin{figure*} [h]
\includegraphics[width=16cm]{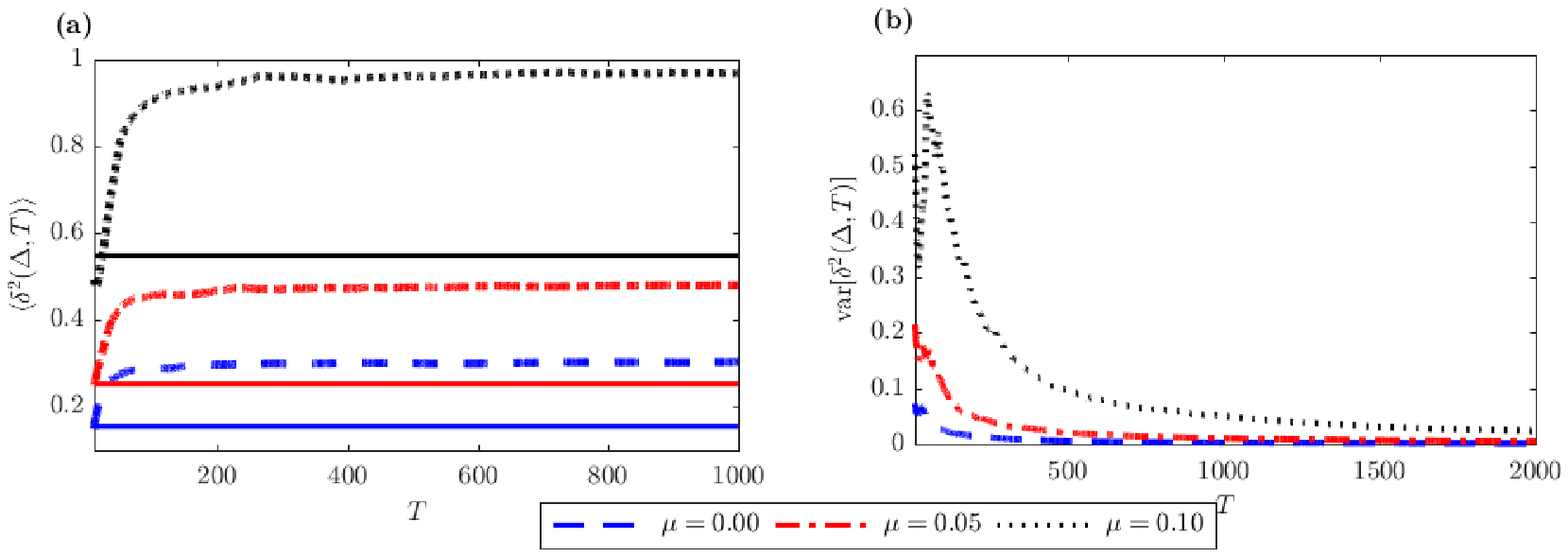} 
\caption{\textbf{TAMSD in diffusion with stochastic resetting}. \textbf{(a)} Mean value of TAMSD as a function of time for various magnitudes of drift $\mu$. We set $r = 0.15$ and $\Delta = 10$. The filled lines are the corresponding MSDs showing deviation from the TAMSD. This is a clear manifestation of ergodicity breaking in TAMSD.
\textbf{(b)} Temporal decay of the variance of TAMSD. Simulation parameters for \textbf{(a-b)}: We set $\sigma^2 = 0.03$ and $dt = 0.01$ and the results are averaged over $10^3$ realizations.  \label{fig:tamsd}}
\end{figure*}

By substituting Eqs.~\eqref{eq:time-average-second-moment} and~\eqref{eq:time-average-autocorrelation} into~\eqref{TAMSD-reset}, we obtain the following expression for the TAMSD in the long time limit
\begin{align}
   \delta^2(\Delta) &= 2 \left(\frac{\mu^2}{r^2} + \frac{\sigma^2}{r} \right) \left( 1 - e^{-r\Delta}\right)\nonumber  \\
  &= \left\langle x^2(\Delta)\right\rangle + \frac{2\mu^2}{r}\Delta e^{-r\Delta} + \frac{\sigma^2}{r}\left(1-e^{-r\Delta}\right),
  \label{eq:long-time-tamsd}
\end{align}
which is a clear violation of \eqref{TAMSD-ergodicity-equivalence}.
In general, the TAMSD is larger than the MSD and, hence, it is a non-ergodic observable in drift-diffusion under stochastic resetting. In Fig.~\ref{fig:tamsd}(a-b) we numerically evaluate the temporal evolution of the TAMSD). The first moment converges to \eqref{eq:long-time-tamsd} and the variance converges to zero, as it should.

\subsubsection{TAMSD and autocorrelations:~}

In order to understand what drives this non-ergodicty, we rewrite the form of the long time TAMSD using \eqref{TAMSD-reset} as
\begin{align}
   \delta^2(\Delta) &= 2 \langle x^2(\Delta) \rangle \left( 1 - \mathrm{corr}\left[\Delta\right] + \frac{\langle x(\Delta) \rangle^2}{\langle x^2(\Delta) \rangle} \right),
\end{align}
which holds for any stationary and ergodic processes. This is a very important result since the TAMSD can be written in terms of the MSD as well as the autocorrelation function (also see \cite{jeon2012inequivalence}). From the above equation it is evident that the autocorrelation essentially drives the magnitude of the TAMSD in diffusion with resetting: lower correlation leads to larger TAMSD and vice versa.

To see this, let us first consider the reset-free simple diffusion process. In such case, the system is non-stationary, and the normalized autocorrelation is given by $\sqrt{\frac{t}{t+\Delta}}$, which saturates to unity at large time. Thus the motion is strongly correlated and the particle traces its path not very different way than its current position lowering the overall TAMSD. On the other hand, by renewing the process through resetting dynamics renders correlation lower than unity and as a result, the particles spread a wider range causing a larger TAMSD. 

We point out a similar discrepancy between the TAMSD and MSD that was found for driftless but reset-FBM ~\cite{biorxiv}. Concretely, the authors found out that for large $\Delta$ the TAMSD is twice the magnitude of the MSD. This is also true in ordinary diffusion with resetting [by setting $\mu = 0$ in \eqref{eq:long-time-tamsd}] and one finds
\begin{align}
\delta^2(\Delta) = 2 \langle x^2( \Delta) \rangle.
\end{align}
However, unlike in FBM, in standard diffusion there is no Hurst exponent which drives the correlations between non-reset trajectories and determines the magnitude of the TAMSD for low values of $\Delta$.  Therefore, in the driftless diffusion case, we always observe this relationship.
For any $\mu$, the generic relationship can be obtained using \eqref{eq:long-time-tamsd}.

\subsubsection{Discontinuity in TAMSD and a critical resetting rate:~}
\begin{figure*}[h]
\includegraphics[width=10cm]{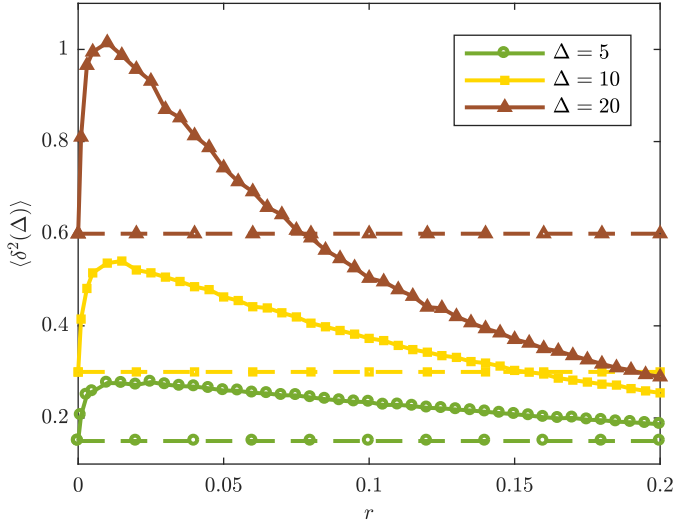} 
\caption{\textbf{Long time TAMSD in diffusion with stochastic resetting}. We plot the average $\langle \delta^2(\Delta) \rangle$ of the long time TAMSD (at $T = 2000$ and averaged across $10^3$ realizations)  as a function of the resetting rate for various lag time $\Delta$. The dashed horizontal lines correspond to $\delta^2(\Delta) = \langle x^2(\Delta)\rangle$ for $r=0$ case.  We set $\mu = 0$, $\sigma^2 = 0.03$ and $dt = 0.01$.
\label{fig:long-time-tamsd}}
\end{figure*}

Interestingly, we observe a discontinuity in the magnitude of the long time TAMSD as the resetting rate approaches zero (see Fig.~\ref{fig:long-time-tamsd} where we plot the numerical average $\langle \delta^2(\Delta) \rangle$ of the long time TAMSD). To delve deeper, let us again consider the simple diffusion case i.e., $\mu=0$ in the absence of resetting ($r=0$) so that we have $\delta^2(\Delta) = \langle x^2(\Delta)\rangle$. This is known as the $r = 0$ case corresponds to the ergodic situation of free Brownian motion, (Fig.~\ref{fig:long-time-tamsd} horizontal dashed lines). But as soon as $r >0^+$, we observe $\delta^2(\Delta) =2 \langle x^2(\Delta) \rangle$ thus breaking the ergodicity immediately. Translating to the drift-diffusion process, one also sees a sudden jump in magnitude $\delta^2(\Delta)-\delta^2(\Delta)_{r = 0}$ as a finite $r>0^+$ is switched on (see Fig.~\ref{fig:long-time-tamsd}).

The TAMSD is not a monotonic function of the resetting rate as can be seen from Fig.~\ref{fig:long-time-tamsd}. It first increases and then monotonically decreases as the resetting rate is increased. At a critical resetting rate, the TAMSD becomes smaller than that of the reset-free case. This critical resetting rate $r^*$ can be calculated simply by setting $\delta^2(\Delta)_{r = 0} = \delta^2(\Delta)|_{r^*}$ which results in the following 
transcendental equation
\begin{align}
  e^{-r^*\Delta} &= \frac{r^* \Delta}{2}\left(\frac{\mu^2 \Delta + \sigma^2}{\frac{\mu^2}{r^*} + \sigma^2}\right).
\end{align}
Thus, any resetting rate below $r^*$ will increase the temporal spread of a single diffusing particle. On the other hand when $r > r^*$, resetting occurs on a far more frequent basis. This hinders the spread of an individual particle thus reducing TAMSD.

\section{Conclusion}
\label{sec:conclusion}
Ergodicity is of paramount importance in physics and economics. Despite many years of rigorous studies in statistical physics, ergodic properties have not been studied much in resetting systems. Here, we took exactly this task and studied ergodic behavior of a diffusing particle in the presence of resetting. To this end, we presented a general formalism that allows us to write explicit solutions to the diffusion process (also with added drift) that is subjected to stochastic resetting. Similar to previous works, our approach is based on the renewal properties of the resetting dynamics but on a single trajectory level. This method is particularly useful since we can use the law of total expectation to quantify the moments and other statistical metrics such as the autocorrelation functions.

Our analysis showed that diffusion with stochastic resetting is \textit{mixing}. This implies that observables such as the sample mean, the squared displacement and the autocorrelation are ergodic. That is, the results will be indistinguishable if we average over an ensemble or along a long observation time in order to obtain the typical behavior of the resetting system. However, the mixing property of the process also indicates that measures which are especially made for tracking the spread in single-particle dynamics such as the TAMSD will be non-ergodic (see also Ref.~\cite{jeon2012inequivalence}).  We observe a pronounced ramification in the magnitude of the TAMSD due to resetting. Interestingly, we found out that there is a critical resetting rate $r^*$: for any rate below it, resetting increases the time-averaged spread of a single particle. On the other hand, for $r>r^*$, we have found that resetting constraints the spread of the particle thus lowering the TAMSD. Finally, we stress that our method is comprehensive since it bridges two different approaches (namely analysis of sample mean in economics and TAMSD in physics) to study ergodic properties in a resetting system. It remains to be seen how the statistical properties of the sample mean and TAMSD alter intricately under arbitrary resetting time density.

Besides this direction, our analysis unraveled a handful of implications that need to be addressed in greater detail in the near future. For example, here we analyzed the TAMSD in the long time limit. For finite times, the TAMSD will be a random variable and there will be an ergodicity breaking parameter that will quantify the extent to which ergodicity is broken \cite{TAMSD-review-1,TAMSD-review-2,akimoto_prl}. In this aspect, it would be interesting to explore its relationship with the resetting rate. It is also worth emphasizing that the explicit solution to diffusion with resetting is general and can be applied to any Markovian process that is subjected to resetting, such as the geometric Brownian motion~\cite{GBM} and the Ornstein-Uhlenbeck process~\cite{Pal-potential,meylahn2015large}. Extending our study to non-Markovian processes such as generalized geometric Brownian motion \cite{stojkoski2020generalised} or anomalous diffusion~\cite{metzler2000random} is a future challenge and will be explored somewhere else. We conclude by stating that our results can be verified in resetting experiments that can track trajectories of a single particle using optical tweezers \cite{expt-1}. Naturally, we anticipate that our work will open a research avenue that will bring synergy between theory and experiment and furthermore will attract various applications in statistical, chemical, and biological physics.

\section*{Acknowledgments}

VS, TS and LK acknowledge financial support by the German Science Foundation (DFG, Grant number ME 1535/12-1). TS was supported by the Alexander von Humboldt Foundation. AP gratefully acknowledges support from the Raymond and Beverly Sackler Post-Doctoral Scholarship and the Ratner Center for Single Molecule Science at Tel-Aviv University.

\appendix

\label{appendix}

\renewcommand{\thesection}{Appendix \arabic{section}}

\section{Mixing and long time relaxation in \\ drift-diffusion with Poissonian resetting}
\label{app:relaxation}

In Section~\ref{mixing-prelim}, we showed how a vanishing correlation function implies mixing properties of a stochastic process of interest. In this section, we present 
another approach based on the $\beta$-mixing coefficient
to study mixing properties of a system. The $\beta(t)$ coefficient is basically the total variational distance between the PDF of the process in time $t$, $P_r(x,t)$ and the stationary distribution of the process $P^*(x)$, i.e.,
\begin{align}\label{beta-mixing}
    \beta(t) &= || P_r(x,t) - P^*(x) ||,
\end{align}
where $||g(y)|| = \int |g(y)| dy$ is the $L^1$ norm of $g$. A stochastic process is said to be $\beta$-mixing if $\lim_{t \to\infty} \beta(t) = 0$ \cite{risteski}. In other words, a system will be mixing if its evolution over time eventually converges to the stationary distribution.

As elaborated in the main text, in a drift-diffusion process subjected to Poissonian resetting, the evolution of the PDF is described with \eqref{renewal eq} and the stationary PDF is given with  \eqref{ness}. Convergence to the stationary state, and thus the evaluation of $\beta(t)$, can be done by adopting a large deviation function. In particular, it is well-known that for the reset-free Brownian motion position distribution is Gaussian and the width grows with time as $\sqrt{t}$. Thus the system does not have any stationary state. When resetting is introduced, a non-equilibrium steady state is established in an inner core region $[-\xi(t),\xi(t)]$ around the resetting point $x_0=0$, where the length scale is $\xi(t)\sim t$. Outside this core region the system is still found to be in transient \cite{review}. At large time, only the second term from the RHS of the renewal equation~(\ref{renewal eq}) is relevant and we proceed with the analysis of its behavior in the following. Substituting the propagator of the drift-diffusion process from \eqref{shifted gaussian} into \eqref{renewal eq} gives us
\begin{align}
    I=r\int_{0}^{t}e^{-rt'}\frac{1}{\sqrt{2\pi \sigma^2 t'}}e^{-\frac{(x-\mu t')^2}{2\sigma^2 t'}}\,dt'.
\end{align}
We make a change of variables $t'=t\tau$ ($dt'=td\tau$) in order to simplify the integral to
\begin{align}
    I=\frac{r}{\sqrt{2\pi \sigma^2}}\sqrt{t}\int_{0}^{1}e^{-rt\tau}e^{-\frac{(x-\mu t\tau)^2}{2\sigma^2t\tau}}\,d\tau=\frac{r}{\sqrt{2\pi \sigma^2}}\sqrt{t}\int_{0}^{1}e^{-t\left[r\tau+\frac{(w-\mu\tau)^2}{2\sigma^2\tau}\right]}\,d\tau,
\end{align}
where $w=x/t$. Moreover, we introduce the function
\begin{align}\label{Phi general}
    \Phi(\tau,w)=r\tau+\frac{(w-\mu\tau)^2}{2\sigma^2\tau},
\end{align}
which further simplifies $I$ as
\begin{align}\label{I final}
    I=\frac{r}{\sqrt{2\pi \sigma^2}}\sqrt{t}\int_{0}^{1}e^{-t\Phi(\tau,w)}\,d\tau.
\end{align}
For calculation of the integral
\begin{align}
    \mathcal{I}(t)=\int_0^1e^{-tf(z)}g(z)dz,
\end{align}
for large $t$, we use the Laplace approximation \cite{arfken} which gives
\begin{align}
    \mathcal{I}(t)\approx e^{-t\,f(z_0)}g(z_0)\sqrt{\frac{2\pi}{t|f''(z_0)|}},
\end{align}
where $z_0$ is the saddle point of the function $f(z)$ such that $f'(z_0)=0$ with $z_0<1$. In case the the saddle point is outside the integration limits ($z_0>1$), then the approximation result should be calculated at $z_0=1$. From (\ref{Phi general}) we can obtain the saddle point $\tau^*$ by setting $\partial_{\tau}\Phi(\tau,w)|_{\tau^*}=0$ which yields
\begin{align}
    \tau_*=\frac{|w|}{\sqrt{\mu^2+2\sigma^2r}}.
\end{align}
Substituting $\tau^*$ into \eqref{Phi general}, we find
\begin{align}\label{Phi general <1}
    \Phi(\tau_*,w)=-\frac{\mu}{\sigma^2}w+\frac{\sqrt{\mu^2+2\sigma^2r}}{\sigma^2}|w|, \quad \text{for} \quad \tau_*<1.
\end{align}
Outside the region, we have
\begin{align}\label{Phi general >1}
    \Phi(1,w)=r+\frac{(w-\mu)^2}{2\sigma^2}, \quad \text{for} \quad \tau_*>1.
\end{align}
Therefore, the PDF behaves as
\begin{align}
    P_{r}(x,t)\sim e^{-t\,I_{r}(x/t)},
\end{align}
where $I_{r}(x/t)$ is the large deviation function given by
\begin{align}\label{ldf2}
    I_{r}(x/t)=\left\lbrace\begin{array}{ll}
         -\frac{\mu}{\sigma^2}\frac{x}{t}+\frac{\sqrt{\mu^2+2\sigma^2r}}{\sigma^2}\frac{|x|}{t}, & |x|<\sqrt{\mu^2+2\sigma^2r}\,t, \\
         r+\frac{(x/t-\mu)^2}{2\sigma^2}, & |x|>\sqrt{\mu^2+2\sigma^2r}\,t.
    \end{array}\right.
\end{align}
This result defines the separation of the region in which relaxation has been achieved ($\tau_*<1$) from those in which the system is still not relaxed---it is in transient state ($\tau_*>1$). The above results tells us that $\beta(t) \sim e^{-(r+\frac{\mu^2}{2\sigma^2})t}$ (see \eqref{beta-mixing} and the convergence criterion of $\beta(t)$). Hence, the $\beta$-coefficient analysis conforms with the findings from the main text that drift-diffusion with stochastic resetting is mixing. For further details of such analysis and simulations of the LDF and the transition to the steady state we refer to \cite{transport1,jpa2020_rk,arxiv2021_rk}. A complementary approach based on the spectral decomposition of the Fokker-Planck equation was taken in \cite{return-relax} to study relaxation properties of diffusion with instantaneous resetting.

{}

\end{document}